\documentclass[two column,superscriptaddress,groupedaddress amssymb, amsmath,nobibnotes,nofootinbib, aps, showpacs, prb]{revtex4-1}
\usepackage[dvips]{graphicx}
\usepackage{graphicx}
\usepackage{color}

\begin{document}

\title{Multigap superconductivity in ThAsFeN investigated using $\mu$SR measurements}
\author{Devashibhai Adroja}
\email{devashibhai.adroja@stfc.ac.uk}
\affiliation{ISIS Facility, Rutherford Appleton Laboratory, Chilton, Didcot Oxon, OX11 0QX, United Kingdom} 
\affiliation{Highly Correlated Matter Research Group, Physics Department, University of Johannesburg, PO Box 524, Auckland Park 2006, South Africa}
\author{Amitava Bhattacharyya}
\email{amitava.bhattacharyya@rkmvu.ac.in}
\affiliation{ISIS Facility, Rutherford Appleton Laboratory, Chilton, Didcot Oxon, OX11 0QX, United Kingdom} 
\affiliation{Department of Physics, Ramakrishna Mission Vivekananda University, Howrah-711202, India}
\author{Pabitra Kumar Biswas}
\affiliation{ISIS Facility, Rutherford Appleton Laboratory, Chilton, Didcot Oxon, OX11 0QX, United Kingdom} 
\author{Michael Smidman}
\affiliation{Center for Correlated Matter and Department of Physics, Zhejiang University, Hangzhou 310058, China}
\author{Adrian Hillier}
\affiliation{ISIS Facility, Rutherford Appleton Laboratory, Chilton, Didcot Oxon, OX11 0QX, United Kingdom} 
 \author{Huican Mao}
\affiliation{Beijing National Laboratory for Condensed Matter Physics, Institute of Physics, Chinese Academy of Sciences, Beijing 100190, China}
\author{Huiqian Luo}
\affiliation{Beijing National Laboratory for Condensed Matter Physics, Institute of Physics, Chinese Academy of Sciences, Beijing 100190, China}
\author{Guang-Han Cao}
\affiliation{Department of Physics and State Key Lab of Silicon Materials, Zhejiang University, Hangzhou 310027, China}
\author{Zhicheng Wang}
\affiliation{Department of Physics and State Key Lab of Silicon Materials, Zhejiang University, Hangzhou 310027, China}
\author{Cao Wang} 
\affiliation{Department of Physics, Shandong University of Technology, Zibo 255049, China}

\date{\today}

\begin{abstract}

We have investigated the superconducting ground state of the newly discovered superconductor ThFeAsN with a tetragonal layered crystal structure  using resistivity, magnetization, heat capacity and transverse-field (TF) muon-spin rotation ($\mu$SR) measurements. Our resistivity and magnetization measurements reveal an onset of bulk superconductivity with $T_{\bf c}\sim$ 30 K. The heat capacity results show a very small anomaly, $\Delta$C$_{ele}$$\sim$0.214 (J/mol-K) at $T_{\bf c}\sim$ 30 K and exhibits exponential behavior below  $T_{\bf c}$, which fits better to two superconducting gaps rather than a single gap.  Further a nonlinear magnetic field dependence of the electronic specific heat coefficient $\gamma$(H) has been found in the low temperature limit, which indicates that the smaller energy gap is nodal.  Our analysis of the TF-$\mu$SR results shows that the temperature dependence of  the superfluid density is better described by a two-gap model either isotropic  $s$+$s$-wave or $s$+$d$-wave than a single gap isotropic $s$-wave model for the superconducting gap, consistent with other Fe-based superconductors.  The combine $\gamma$(H) and TF-$\mu$SR results confirm $s$+$d$-wave model for the gap structure of ThFeAsN.  The observation of two gaps in ThFeAsN suggests multiband nature of the superconductivity possibly arising from the d-bands of Fe ions.  Furthermore,  from our TF-$\mu$SR study we have estimated the magnetic penetration depth, in the polycrystalline sample, of $\lambda_{\mathrm{L}}$$(0)$ = 375 nm, superconducting carrier density $n_s = 4.6 \times 10^{27}~ $m$^{-3}$, and carrier's effective-mass $m^*$ = 2.205\textit{m}$_{e}$. We will compare the results of our present study with those reported for the Fe-pnictide family of superconductors.
\end{abstract}


\maketitle

\section{Introduction}

In a conventional superconductor, the binding of electrons into the paired states, known as the Cooper pairs,  that collectively carry the supercurrent is mediated by lattice vibrations  or so called phonons. This is the fundamental  principle of the Bardeen-Cooper-Schrieffer (BCS) theory~\cite{B.C.S.}. However, the BCS theory often fails to describe the superconductivity (SC) observed in strongly correlated materials. Several strongly correlated superconducting materials, having magnetic  \textit{f}$-$ or \textit{d}$-$ electron  elements, exhibit unconventional SC and various theoretical models based on magnetic interactions (magnetic glue)  and spin fluctuations have been proposed to understand these superconductors~\cite{U.C.S., Stewart2017}. Gauge symmetry is broken in the case of conventional BCS superconductors and other symmetries of the Hamiltonian are broken for unconventional superconductors in the superconducting state. BCS superconductors can also show gap anisotropy, although they remain nodeless and the gap does not change sign over the Fermi surface, while  unconventional superconductors may have nodes (zeros) in the gap function along certain directions and the location of the nodes is closely associated with the pairing symmetry. Therefore investigation of the superconducting gap structure of strongly correlated \textit{f}$-$ and \textit{d}$-$ electron superconductors is very important for understanding the physics of unconventional pairing mechanisms in these classes of materials.

\par

Unconventional superconductivity has been  observed in  high-temperature cuprates ~\cite{H.T.S.C.}, iron pnictides ~\cite{FeAs} and heavy fermion materials~\cite{H.F.S.C.}, which  have strong electronic correlations and quasi-two-dimensionality. Interestingly superconductivity in the iron based materials emerges after doping electrons/holes into an antiferromagnetic parent compound~\cite{FeAs}, for example, LaFeAsO$_{1-x}$F$_x$(1111 family)~\cite{de,Huang}, BaFe$_{2-��x}$Co$_x$As$_2$ (122 family) ~\cite{Shibauchi}, NaFe$_{1-��x}$Co$_x$As (111 family)~\cite{Parker}, FeTe$_{1-��x}$Se$_x$ (11 family) ~\cite{Liu,Katayama} and Ca$_{1-��x}$La$_x$FeAs$_2$ (112 family) ~\cite{Katayama1,Jiang} etc. Some special systems are self-doped by the ion deficiency, such as LaFeAsO$_{1-��\delta}$~\cite{Ren} and Li$_{1-��\delta}$FeAs~\cite{Wang}. It is interesting that in the 1111 family of Fe-based materials, superconductivity can be induced by chemical substitution (i.e. electron and hole-doping) on any atomic site, for example an antiferromagnetically ordered ground state in LaFeAsO is transformed into a superconducting ground state with  fluorine  and hydride doping on the oxygen site (e.g., LaFeAsO$_{1-��x}$F$_x$ LaFeAsO$_{1-��x}$H$_x$)~\cite{Matsuishi,Zhu,Iimura,Hiraishi}.

\par

It is of great interest to explore possible unconventional superconductivity in stoichiometric Fe-based layers materials, having a tetragonal crystal structure, with significant electron correlations. Recently, the first nitride iron pnictide superconductor ThFeAsN, containing layers with nominal compositions [Th$_2$N$_2$] and [Fe$_2$As$_2$] (inset of Fig.1), has been discovered, with $T_c$ = 30 K for the nominally undoped compound ~\cite{CWang2016}. The transition temperature of this newly discovered material is as high as electron-doped 111-based superconductors and another newly discovered stoichiometric superconductors ACa$_2$Fe$_4$As$_4$F$_2$ (A=K, Rb and Cs, $T_c$$\sim$ 30 K) ~\cite {ZCWang2016, Zhi-ChengWang2017}. Although the first-principle calculations of ThFeAsN indicates the lowest energy magnetic ground state is the stripe-type antiferromagnetic state ~\cite {DSing, GWang2016}, the normal state resistivity shows no obvious magnetic anomaly, but only metallic behavior down to 30 K~\cite {CWang2016}. The self-doping effect, which arises due to covalent N-N bonding that reduces the effective nitrogen valence, may be responsible for the superconductivity and complete suppression of the magnetic order in this material. Any further electron doping by substituting N with O or hole doping by substituting Th with Y only suppress the superconducting $T_c$. The DFT calculations of ThFeAsN shows  approximately nested hole and electron Fermi surfaces of Fe d character involving the xz; yz and xy orbitals, indicating strong similarity to the other Fe-pnictide families of superconductors ~\cite {DSing, GWang2016}. ThFeAsN shares similar electronic structure and magnetic properties to those of LaOFeAs ~\cite {GWang2016}. The calculated bare susceptibility $\chi$$_{0}$(q) of ThFeAsN peaks at the M-point, suggesting perfect nesting between the hole-like and electron-like fermi surface with vector {\bf q} = ($\pi$,  $\pi$, 0), similarly to other FeAs-based superconductors~\cite {GWang2016}. Further, the non-magnetic ground state of ThFeAsN down to 2 K has been confirmed through neutron powder diffraction measurements~\cite{HMao2017} and a $^{57}$Fe M$\ddot{\mathrm{o}}$ssbauer spectroscopy study on polycrystalline samples~\cite{Moss}.

\par

In addition to all the existing information, it is important to understand the superconducting and magnetic properties of ThFeAsN in a microscopic level. The transverse field (TF) muon spin rotation and relaxation ($\mu$SR) measurements provide direct information on the nature of the superconducting gap symmetry and absolute value of the magnetic penetration depth.  We therefore have investigated the superconducting properties of ThFeAsN using the bulk properties and TF-$\mu$SR measurements. Our study of the TF-$\mu$SR shows that the temperature dependence of  the superfluid density is better described by a two-gap isotropic  $s$+$s$-wave model than a single gap isotropic $s$-wave model.

\begin{figure}[t]
\centering
\includegraphics[width =  \linewidth,trim={-30mm -10mm 0 0mm},clip ]{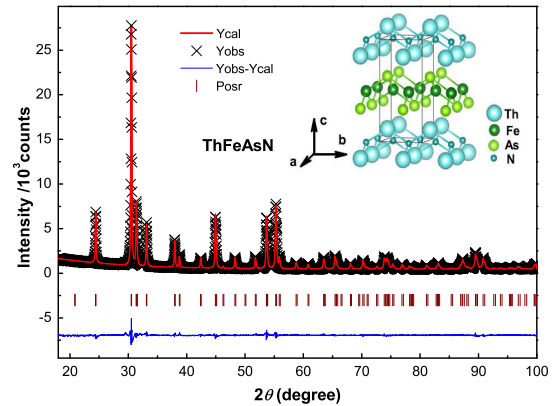}
\caption {(Color online) X-ray powder-diffraction pattern (at 300K) with the Rietveld refinement fit of the data of the ThFeAsN  compound. The line drawn through the data points corresponds to the calculated pattern and the cross-symbols represent observed data. The vertical bars show the Bragg peaks' positions and the blue line at bottom shows the difference plot. The inset shows the tetragonal crystal structure.~\cite {HMao2017}}
\label{xrd:fig1}
\end{figure}

\section{Experimental Details}

A polycrystalline sample of ThFeAsN was synthesized by the solid state reaction method as described by Wang {\it et al.}~\cite {CWang2016}. The sample was characterized using powder x-ray diffraction (XRD), electrical resistivity, magnetic susceptibility and heat capacity measurements. The resistivity and heat capacity were measured using a Quantum Design Physical Property Measurement System (PPMS) between 1.5 and 300 K.  Temperature dependent resistivity from 2 K to 300 K was measured by standard 4-probe method. The heat capacity was measured by standard thermal relax method with a sample $m=$ 18 mg. The DC magnetization of the same sample was measured on a Quantum Design Magnetic Property Measurement System (MPMS).Muon spin relaxation/rotation ($\mu$SR) experiments were carried out on the MUSR spectrometer at the ISIS pulsed muon source of the Rutherford Appleton Laboratory, UK~\cite{sll}. The $\mu$SR measurements were performed in  transverse$-$field (TF) mode. A pellet (12mm diameter) of polycrystalline sample ThFeAsN was mounted on a silver  (99.999\%) sample holder.  Hematite ($\alpha$-Fe$_2$O$_3$) slabs were placed just after the sample to reduce the background signal. The sample was cooled under He-exchange gas in a He-4 cryostat operating in the temperature range of 1.5 K$-$300 K.  F$-\mu$SR experiments were performed in the superconducting mixed state in an applied field of 400 G, well above the lower critical field of $H_{c1}$ $\sim$ 30 G of this material.~Data were collected in the field$-$cooled (FC) mode, where the magnetic field was applied above the superconducting transition temperature and the sample was then cooled down to base temperature. Muon spin rotation and relaxation is a dynamic method that allows one to resolve the nature of the pairing symmetry in superconductors~\cite{js}. The vortex state in the case of type-II superconductors gives rise to a spatial distribution of local magnetic fields; which demonstrates itself in the $\mu$SR signal through a relaxation of the muon polarization. The data were analyzed using the free software package WiMDA~\cite{FPW}.

\begin{figure}[t]
\vskip -0.0 cm
\centering
\includegraphics[width= \linewidth, trim={0mm -15mm 0mm 0mm},clip]{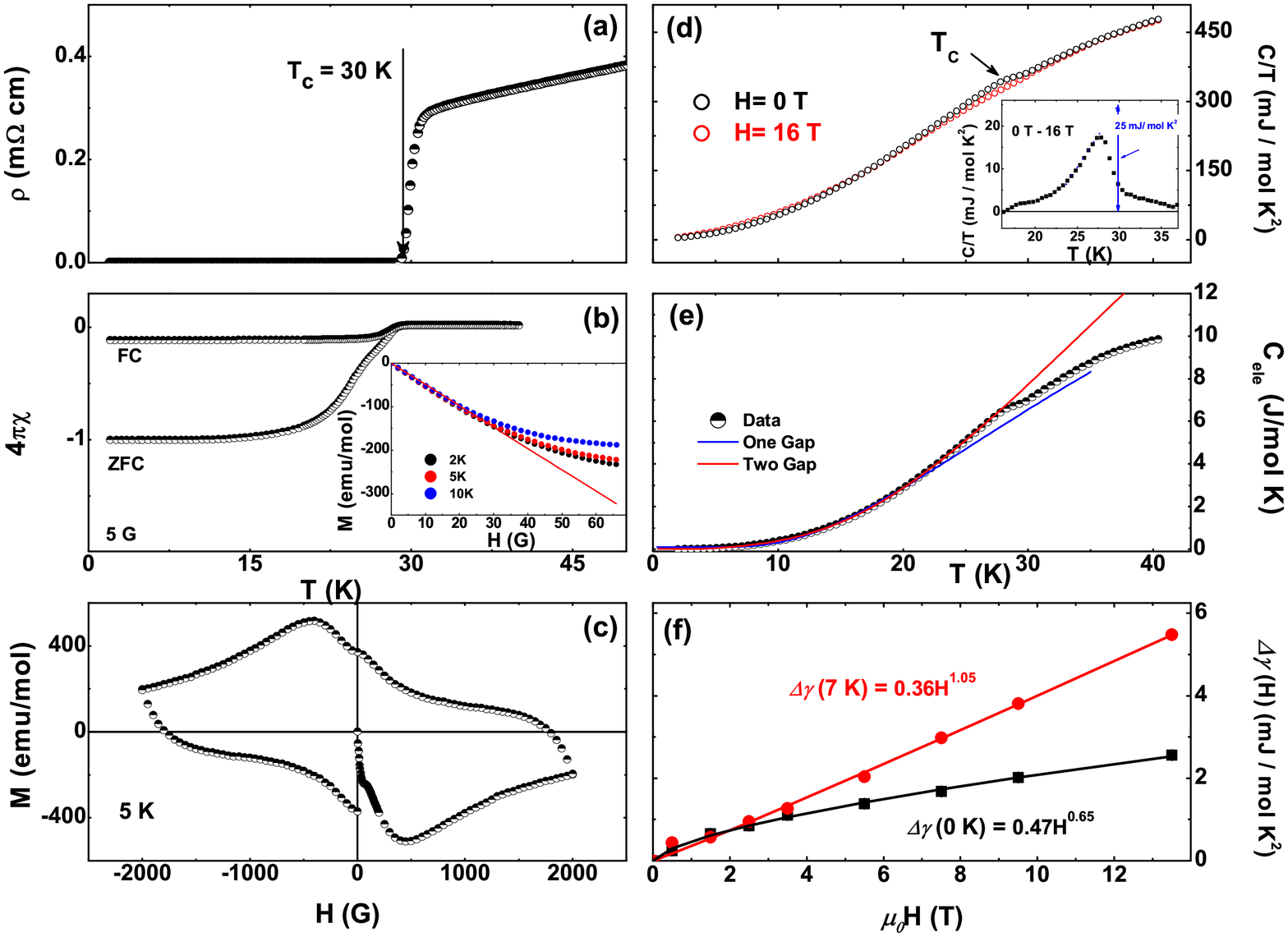}

\caption {(Color online) (a) Temperature dependence of electrical resistivity, (b) Low-field dc-magnetic susceptibility measured in zero-field cooled (ZFC) and field cooled (FC) modes in an applied field of 5 G. Inset shows the isothermal filed dependence of magnetization at 2, 5 and 10 K.
(c) The isothermal filed dependence of magnetization at 5 K.  (d) Temperature dependence of heat capacity in zero field and in applied field of 16 T and the inset shows difference of the heat capacity data, 0 T- 16T, plotted at $C/T$ vs $T$. The blue vertical arrow shows the jump in $C/T$ at $T_C$. (e) The electronic contribution  to the heat capacity. The blue and red lines represent the fit using a single gap and two-gap models. (f) The magnetic field dependence of the electronic specific heat coefficient $\Delta$$\gamma$ (=$\gamma$(H)-$\gamma$(0)) at extrapolated to T$\sim$0 K and 7 K. The solid red line shows a power law fit, $\gamma$(H)$\sim$H$^{n}$.}
\label{bulk:fig2}
\end{figure}

\section{Results and discussions}

The analysis of the powder x-ray diffraction at 300 K reveals that the sample is single phase and crystallizes in the ZrCuSiAs-type tetragonal crystal structure with space group $P4/nmm$ (No. 129, Z = 2) as shown in the inset of Fig. 1. The refined values of the lattice parameters are   a = 4.0367(2) \AA~and c = 8.5262(2) \AA. The  layered structure of ThFeAsN is shown in the inset Fig.1 perpendicular to the c-axis,  where separated layers of Th and N ions at the bottom and top of the unit cell (along the c-axis) can be seen. The As and Fe layers are at half way along the c-axis. The Fe and As ions form tetrahedrons with two As-Fe-As bond angles  $\alpha$$\sim$107.0$^\circ$ and $\beta$$\sim$114.5$^\circ$ at 300 K. The layered structure of ThFeAsN is very similar to other 1111 family of iron-pnictide superconductors~\cite{Zhi-An Ren}. 

\par

The electrical resistivity reveals  a sharp drop below 30 K  followed by zero-resistivity indicating the onset of superconductivity with $T_c$ = 30 K. In zero field, the temperature dependent resistivity of ThFeAsN is metallic and exhibits a power low behavior $\rho$ = $\rho$$_{0}$+a*T$^{n}$ with n$\sim$ 1.3 between $T_c$ and 150 K, indicating non-Fermi-liquid behavior~\cite {HMao2017}. The low-field magnetic susceptibility measured in an applied field of 5 G shows an onset of diamagnetism below 30 K indicating that the superconductivity occurs at 30 K and the superconducting volume fraction is close to 100\% at 5.0 K  [Fig. 2(b)]. This result confirms the bulk nature of superconductivity with $T_c$ = 30 K in ThFeAsN, which is comparable to $T_c$ = 26 K observed in fluorine-doped LaFeAsO~\cite {LaFeAsOF}. Very similar behavior of the resistivity and magnetic susceptibility has been reported for ThFeAsN  by  H. Mao {\it et al.}~\cite {HMao2017}.  

\par

The magnetization isotherm $M\left(H\right)$ curve at 5 K [inset of Fig. 2(b)] shows typical behaviour for type-II superconductivity. The lower critical field $H_{c1}$ obtained from the M vs H plot at 5 K is 30 G. The upper critical field $H_{c2}$ = 80 kG~\ at 26 K (with slope dH/dT$\sim$-2.4T/K) has been estimated using field dependent resistivity measurements ~\cite{GCao} compared to the  Pauli limit, $\mu_{0}H_{P} = 18.4T_{\mathrm{c}} = 552$~ kG (55.2 T)~\cite{CAM}.  At the superconducting transition, the jump in $(C/T)$ at $T_c$, $\Delta($C/$T_{\mathrm{c}}$) = 25 (mJ/mol K$^2$), which  is a factor 2.78 larger than 9 (mJ/mol K$^2$) observed in LaFeAsO and SmFeAsO polycrystalline samples ~\cite {Gang2008}. The dimensionless specific heat jump is  $\Delta$C/$\gamma$$T_{\mathrm{c}}$ =0.031, which is smaller than the simple $\it {s}-$wave BCS prediction 1.43~\cite{Z. Tang} and also  smaller than 2.2 observed in K$_2$Cr$_3$As$_3$~\cite{T.Kong}. On the other hand the heavy fermion superconductors exhibit very large jumps in the heat capacity at $T_c$. For example, $\Delta$C/$\gamma$$T_{\mathrm{c}}$ values are 4.5  in CeCoIn$_5$ and 2.7 UBe$_{13}$ and 5.7 (at 2.58 GPa) in CeIrSi$_3$~\cite{CeCoIn5,UBe13,CeIrSi3}. We have estimated the electronic  contribution (C$_{ele}$) to the heat capacity below $T_c$ by estimating the phonon contribution above $T_c$ and subtracting from the total heat capacity. We have fitted the heat capacity data above $T_c$ to  C$_p$ = $\gamma$T + $\beta$T$^3$. The fits yield, $\gamma$ = 0.248 (mJ/(mol-K$^2$) and   $\beta$ =  1.44 $\times$ 10$^{-4}$ (mJ/(mol-K$^4$). Using this value of  $\beta$ we have estimated the Debye temperature, $\Theta$$_D$= 425 K, which is comparable to $\Theta$$_D$ = 280 K for LaFeAsO~\cite{JDong} and 325 K for LaFeAsO$_{0.89}$F$_{0.11}$~\cite {Sefat2008}. The electronic heat capacity below $T_c$ was analyzed using one gap and two gaps models. The estimated value of the gaps are: $\Delta$$(0)$ = 4.36 meV for the one gap model and $\Delta_1$$(0)$ = 6.38 meV and $\Delta_2$$(0)$ = 1.98 meV for the two gaps model. The values of $\chi$$^{2}$ are 0.997 for one gap and 0.993 for two gaps models, which indicates that the two gaps model fits better to the experimental data. This finding is in agreement with our $\mu$SR analysis discussed below. To shade light on the nature of gap symmetry we have carried out field dependent heat capacity measurements up to 16 T field. The $\gamma$ (H) was estimated by plotting $C/T$ vs $T$$^2$ and extrapolating to T$\sim$0 K. The  estimated $\gamma$ (H) at T$\sim$0 K exhibits a nonlinear magnetic field dependence, but at 7 K it shows linear field dependence. The non-linear behavior,  $\gamma$ (H)$\sim$H$^{0.65}$,  found in the low temperature limit indicates the nodal behavior of the smaller energy gap. A very similar behavior of $\gamma$ (H)$\sim$H$^{0.5}$ has been observed in LaFeAsO$_{0.9}$F$_{0.1}$ by Gang {\it et al.,}~\cite {Gang2008} that has been attributed to the nodal gap structure.

\begin{figure}[t]
\centering
\includegraphics[width = \linewidth,trim={-10mm -25mm 0mm 0mm},clip]{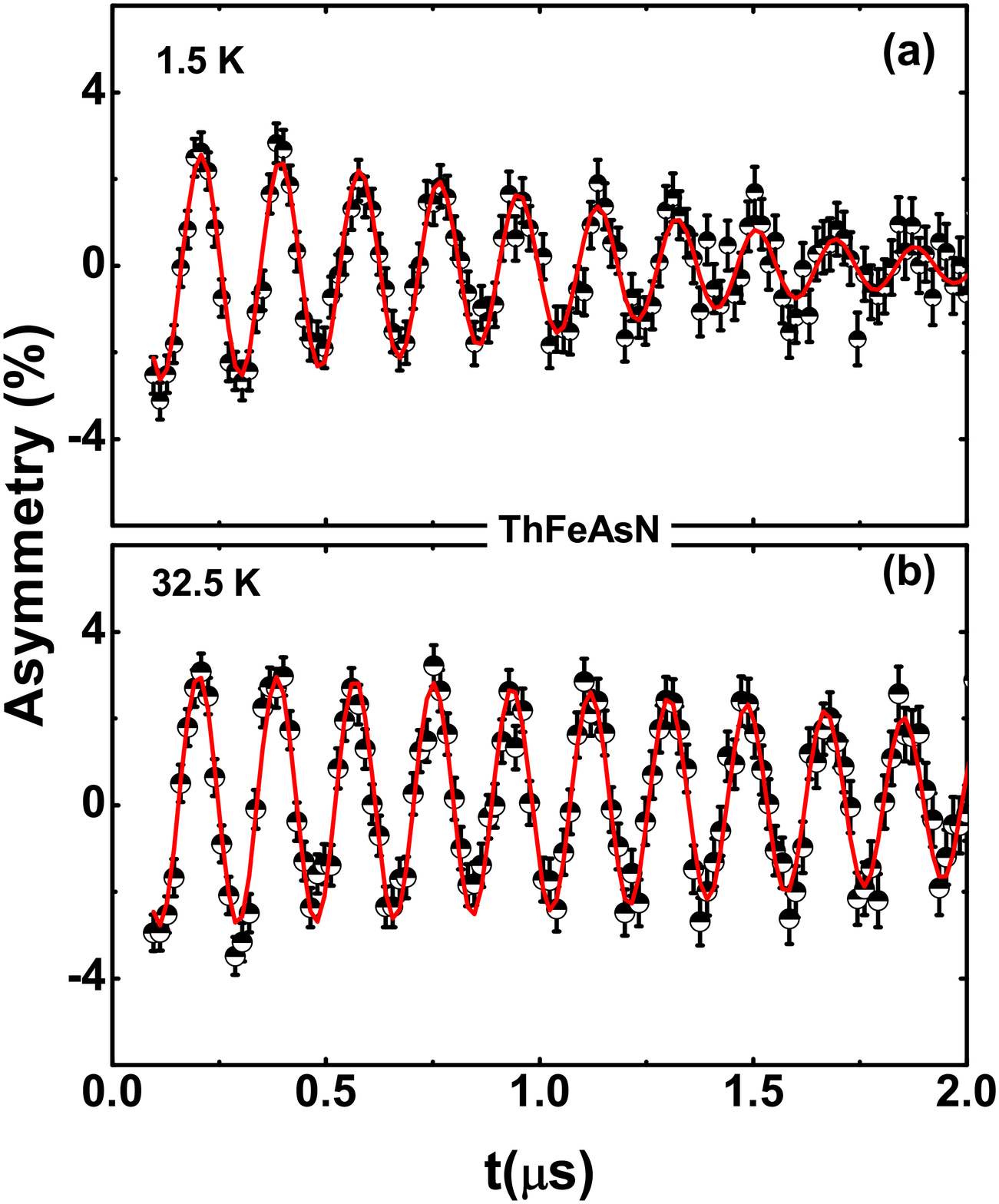}
\caption {(Color online) Transverse field  $\mu$SR  asymmetry spectra for ThFeAsN collected (a) at $T$ = 1.5 K and (b) at $T$ = 32.5  K (i.e. below and above $T_{\bf c}$) at an applied magnetic field of $H$ = 400 G. For the sake of clarity we present here the time-dependent asymmetry in the low time region. The solid line shows a fit using Eq.(1).}
\label{musr1:fig3}
\end{figure} 

\par

Figures 3 (a) and (b) show the TF$-\mu$SR precession signals above and below $T_{\bf c}$ obtained in FC mode with an applied field of 400 G (well above $H_{c1}\sim$ 30 G but below $H_{c2}\sim$ 80 kG, at 26 K). The observed decay of the $\mu$SR signal with time below $T_{\bf c}$ is due to the inhomogeneous field distribution of the flux-line lattice. We have used an oscillatory decaying Gaussian function to fit the TF$-\mu$SR time dependent asymmetry spectra, which is given below,

\begin{equation}
\begin{split}
G_{z1}(t) = A_1\rm{cos}(2\pi \nu_1 t+\phi_1)\rm{exp}\left({\frac{-\sigma^2t^2}{2}}\right)\\ 
\end{split}
\end{equation}

\noindent where $A_1$ is the muon initial asymmetry and $\nu_1$ is the frequency of the muon precession signal associated with the full volume of the sample.  The frequency associated with muon precession on the Hematite (on which the sample pellet was mounted) is very high (209 MHz or 15.48 kG) and is out of the time window of the MUSR spectrometer at the ISIS facility~\cite {Hematite}. Further the both frequency and relaxation rate of Hematite are temperature independent below 100 K~\cite {Hematite}.  In Eq.~(1) the first term contains the total relaxation rate $\sigma$ from the superconducting fraction of the sample; there are contributions from the vortex lattice ($\sigma_{sc}$) and nuclear dipole moments   ($\sigma_{nm}$), where the latter is assumed to be constant over the entire temperature range  [where $\sigma$ = $\sqrt{(\sigma_{sc}^2+\sigma_{nm}^2)}$]. The contribution from the vortex lattice, $\sigma_{sc}$, was determined by quadratically subtracting the background nuclear dipolar relaxation rate obtained from the spectra measured above $\it {T}_{\bf c}$. As $\sigma_{sc}$ is directly related to the superfluid density, it can be modeled by \cite{Prozorov,bhattacharyya2015broken,bhattacharyya2015unconventional}

\begin{equation}
\frac{\sigma_{sc}(T)}{\sigma_{sc}(0)} = 1 + 2 \left\langle\int_{\Delta_k}^{\infty}\frac{\partial f}{\partial E}\frac{E{\rm d}E}{\sqrt{E^2-\Delta_k^2}}\right\rangle_{\rm FS},
\label{RhoS}
\end{equation}

\noindent where $f=\left[1+\exp\left(-E/k_{\mathrm{B}}T\right)\right]^{-1}$ is the Fermi function and the brackets correspond to an average over the Fermi surface. The gap is given by $\Delta(T, \varphi)$=$\Delta_0 \delta(T/\it {T}_c)g(\varphi)$, whereas $g(\varphi)$ refer to the angular dependence of the superconducting gap function and $\varphi$ is the azimuthal angle along the Fermi surface.   We have used the BCS  formula for the temperature dependence  of the gap, which is given by $\delta(T/T_c)$ =tanh$[(1.82){(1.018(\it {T}_c/T-1))}^{0.51}]$~\cite{UBe131}. $g(\varphi)$~\cite{Annett,Pang} is substituted by (a) 1 for $s-$wave gap [also for $s+s$ wave gap or multigap function], (b) $|cos(2\varphi)|$  for $d-$wave gap with line nodes and (c) for anisotropic $s-$wave model $\frac{|1+ cos(4\varphi)|}{2}$~\cite{Prozorov,UBe131,chia,bis2}.

\begin{figure}[t]
\vskip -1.0 cm
\centering
\includegraphics[width = \linewidth,trim={0mm -35mm 0mm 0mm},clip]{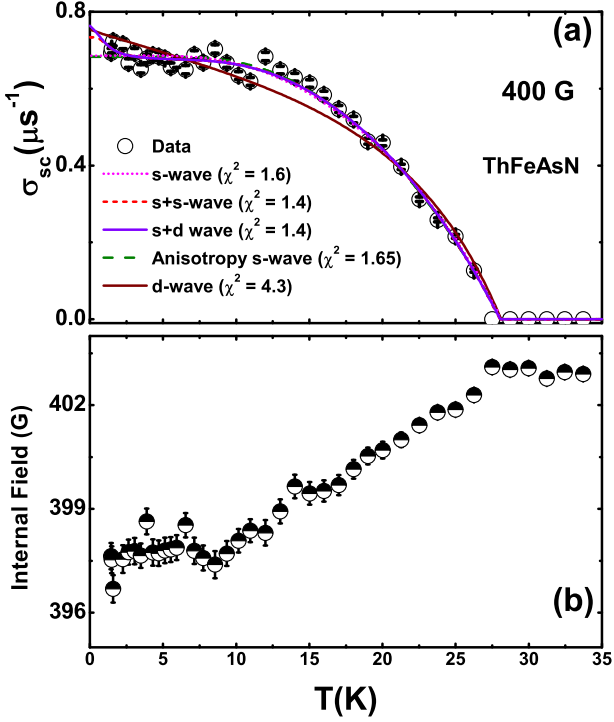}
\caption {(Color online) (a) Temperature dependence of  the muon depolarization rate $\sigma_{sc}(T)$ of ThFeAsN collected in an applied magnetic field of 400 G in field cooled (FC) mode. (b) $\sigma_{sc}(T)$ of FC mode (symbols), where the  lines are the fits to the data using  Eq. 2 for various gap models. The dotted magenta line shows the fit using an isotropic single-gap $s$-wave model with $\Delta(0)/k_{\mathrm{B}}T_{\mathrm{c}}$ = 5.1$\pm0.01$, the dashed red line and blue  solid line show the fit to a two-gap model, $s$+$s$-wave and $s$+$d$-wave, respectively with  $\Delta_1(0)/k_{\mathrm{B}}T_{\mathrm{c}} = 5.2\pm 0.1$ and  $\Delta_2(0)/k_{\mathrm{B}}T_{\mathrm{c}} = 0.3\pm 0.1$ (for both the models). The green long-dashed line shows the fit using an anisotropic $s$-wave model and the solid purple line shows the fit using d-wave model. (b) Temperature dependence of the internal field.}
\label{musr2:fig4}
\end{figure}

\par

Figure 4 (a) shows the temperature dependence of $\sigma_{sc}$, measured in an applied field of 400 G collected in FC mode. The FC mode is thermodynamically stable and provides direct information on the nature of the flux-line lattice. The temperature dependence of  $\sigma_{sc}$ increases with decreasing temperature confirming the presence of a flux-line lattice and indicates a decrease of the magnetic penetration depth ($\lambda^2 \sim \frac{1}{\sigma_{sc}}$) with decreasing temperature. The onset of diamagnetism below the superconducting transition can be seen through the decrease in the internal field below $T_c$ as shown in Fig. 4(b). From the analysis of the observed temperature dependence of $\sigma_{sc}$, using different models for the gap, the nature of the superconducting gap can be determined. We have analyzed the temperature dependence of  $\sigma_{sc}$ based on four different models for the superconducting gap: an isotropic $s$-wave gap model, an isotropic $s$+$s$-wave two-gap model, anisotropic $s$-wave model and a $d$-wave line nodes model. We have also fitted the data using a $s$+$d$-wave two-gap model. The fits to the $\sigma_{sc}(T)$ data of ThFeAsN with various gap models  using Eq. (2) are shown by lines (dashed, dotted and solid) in Fig. 4(a) and the estimated fit parameters are given in Table. I. It is clear from Fig. 4(a) that the $d$-wave model does not fit the data. On the other hand the isotropic $s$-wave, $s$+$s$-wave, $s$+$d$-wave and anisotropic $s$-wave models show good fits to the $\sigma_{sc}(T)$ data. However, upon examining the agreement with the low temperature upturn in the data, it is clear that only two models which explain this feature are  the isotropic $s$+$s$-wave and $s$+$d$-wave two-gap models. Further support of the $s$+$s$- and $s$+$d$-wave models can be seen through goodness of the fit $\chi^2$ given in Table. I. The value of  $\chi^2$ = 1.4 for these models is the lowest. The estimated parameters for the $s$+$s$- and $s$+$d$-wave models show one larger gap $\Delta_1(0)$ = 5.2$\pm$1 (meV) and another much smaller gap $\Delta_2(0)$ = 0.3$\pm$1 (meV). The smaller gap is a nodal gap for $s$+$d$-wave model. Our $\mu$SR analysis alone cannot distinguished between the $s$+$s$- and $s$+$d$-wave models,  but combining the results of field dependent heat capacity, we conclude that the $s$+$d$-wave model is the best to explain observed behavior of $\sigma_{sc}(T)$ and $\gamma$(H). The value of the $\sigma$$_{sc}$(0) =  0.7637$\pm$3~$\mu$s$^{-1}$ and $T_c$ = 28.1$\pm$1 K were estimated from the $s$+$d$-wave fit. The estimated value of 2$\Delta_1(0)$/$k_B$$\it{T}_{\bf c}$ = 4.29$\pm$0.2 from the $s$+$s$- and $s$+$d$-wave fit is comparable to that of the s-wave model (4.21), but larger than the value 3.53 as expected for BCS superconductors. On the other hand for the smaller gap the value 0.3$\pm$0.1 is much smaller than the BCS value. The two-gap nature, one larger and another smaller than the BCS value, are commonly observed in Fe-based superconductors~\cite{twogaps} as well as in Bi$_{4}$O$_{4}$S$_{3}$~\cite{Biswas2013}.  The multigap and $d$-wave order parameters are universal and intrinsic to cuprate superconductors~\cite{Khasanov2007, dwave}, while Cr-based superconductors, A$_2$Cr$_3$As$_3$ (A = K and C) exhibit a nodal gap ~\cite{DTA1,DTA2}. Furthermore, the large value of 2$\Delta_0/k_{\mathrm{B}}T_{\mathrm{c}} = 4.29\pm0.2$ indicates the presence of strong coupling and unconventional superconductivity in ThFeAsN. The two  superconducting gaps (one larger and another smaller) were also observed in SrFe$_{1.85}$Co$_{0.15}$As$_2$,  with $T_c$ = 19.2 K in the STM study~\cite{STMtwogaps}. Moreover combined ARPES and $\mu$SR studies on Ba$_{1-x}$K$_{x}$Fe$_{2}$As$_{2}$ with $T_c$ = 32.0 K also reveal the presence of two gaps ($\Delta$$_1$ = 9.1 meV and $\Delta$$_2$ = 1.5 meV)~\cite{Khasanov2009}.

\begin{figure}[t]
\centering
\includegraphics[width = \linewidth]{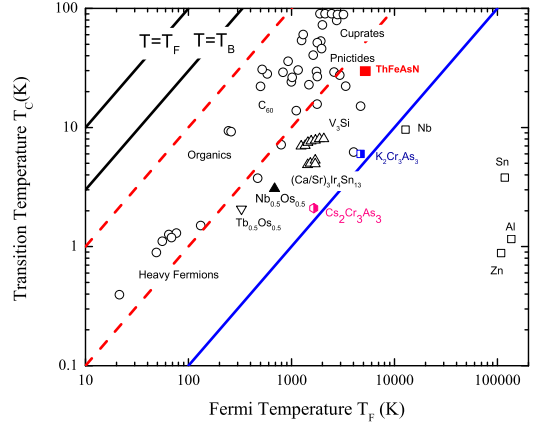}
\caption {(Color online) A schematic representation of the Uemura plot of superconducting transition temperature $T_c$  against effective Fermi temperature $T_F$. The ``exotic'' superconductors fall within a common band for which 1/100$<T_c$/$T_F<$1/10, indicated by the region between two red color dashed lines in the figure. The solid black line correspond to the Bose-Einstein condensation temperature ($T_B$).~\cite{A1}}
\end{figure}

\par

The  muon spin depolarization rate ($\sigma_{sc}$) below $T_{\bf c}$  is related to the magnetic penetration depth ($\lambda$). For a triangular lattice,~\cite{jes,amato,chia} $\frac{\sigma_{sc}(T)^2}{\gamma_\mu^2}= \frac{0.00371\phi_0^2}{\lambda^4(T)}$, where $\gamma_\mu/2\pi$ = 135.5 MHz/T is the muon gyromagnetic ratio and $\phi_0$ = 2.07$\times$10$^{-15}$ T m$^2$ is the flux quantum. This relation between $\sigma_{sc}$ and $\lambda$ is valid for 0.13/$\kappa^{2}$$<<$(H/H$_{c2}$)$<<$1, where $\kappa$=$\lambda$/$\xi$$\gg$70~\cite{Brandt}. As with other phenomenological parameters characterizing a superconducting state, the penetration depth can also be related to microscopic quantities. Using London theory~\cite{jes}, $\lambda_L^2= m^{*}c^2/4\pi n_s e^2$, where $m^* = (1+\lambda_{e-ph})m_e$ is the effective mass and $n_s$ is the density of superconducting carriers. Within this simple picture, $\lambda_L$ is independent of magnetic field. $\lambda_{e-ph}$ is the electron-phonon coupling constant, which can be estimated from $\Theta_D$ and $T_{\mathrm{c}}$ using McMillan's relation~\cite{mcm} $\lambda_{e-ph}=\frac{1.04+\mu^*\ln(\Theta_D/1.45T_{\bf c})}{(1-0.62\mu^*)\ln(\Theta_D/1.45T_{\bf c})+1.04}$, where $\mu^*$ is the repulsive screened Coulomb parameter and usually assigned as $\mu^*$ = 0.13.   For ThFeAsN, we have used $T_{\bf c}$ = 28.1 K and $\Theta_D$ = 425 K , which together with $\mu^*$ = 0.13, we have estimated $\lambda_{e-ph}$ = 1.205. Further assuming that roughly all the normal state carriers ($n_e$) contribute to the superconductivity (i.e., $n_s\approx n_e$),   we have estimated the magnetic penetration depth $\lambda$, superconducting carrier density $n_s$, and effective-mass enhancement $m^*$  to be $\lambda_L(0)$ = 375 nm (from the s+d wave fit), $n_s$ = 4.6$\times$10$^{27}$ carriers/m$^3$, and $m^*$ = 2.205$m_e$, respectively.  

\par

\begin{table}[b]
\begin{center}
\caption{Fitted parameters obtained from the fit to the $\sigma_{sc}(T)$ data of ThFeAsN using different gap models.\\}
\begin{tabular}{lccccccccccccc}
\hline
\hline
 Model && $g(\phi)$ &&Gap value &&  Gap ratio   & $\chi^2$\\ 
&& &&$\Delta(0)$ (meV) &&  2$\Delta(0)/k_B T_{\bf c}$   &\\ 
\hline
$s$ wave && 1 && 5.1(1) &&  4.21   & 1.6\\ 
$s$+$s$ wave&&  1 && 5.2(1); 0.3(1) &&  4.29; 0.3  & 1.4\\ 
anisotropy gap && $\frac{|1+ cos(4\phi)|}{2}$&& 6.29 &&  5.2   & 1.63\\ 
$d$ wave && $\small{cos(2\phi)}$ && 7.75 &&  6.40  & 4.3\\ 
$s$+$d$ wave && 1, $\small{cos(2\phi)}$ && 5.2(1); 0.3 &&  4.29; 0.3  & 1.4\\ 

\hline 
\end{tabular}
\end{center}
\end{table}

\par
The correlation between $T_c$ and $\sigma_{sc}$ observed in $\mu$SR studies has suggested a
new empirical framework for classifying superconducting materials~\cite{U1}. Here we explore the role of muon spin relaxation rate/penetration depth in the superconducting state for the characterisation and classification of superconducting materials as first proposed by Uemura {\it et al.}~\cite{U1}.  In particular we focus upon the Uemura classification scheme which considers the correlation between the superconducting transition temperature, $T_c$, and the effective Fermi temperature, $T_F$, determined from $\mu$SR measurements of the penetration depth~\cite{A1}. Within this scheme strongly correlated ``exotic'' superconductors, i.e. high $T_c$ cuprates, heavy fermions, Chevrel phases and the organic superconductors, form a common but distinct group, characterised by a universal scaling of $T_c$ with $T_F$ such that 1/10$>(T_C/T_F)>$1/100 (Fig. 5). For conventional BCS superconductors 1/1000$>(T_c/T_F$). Considering the value of $T_c/T_F$ = 30/4969.4 = 0.006 for ThFeAsN (see Fig. 5), this material can be classified as not an exotic superconductor, but very close to this limit,  according to Uemura's classification~\cite{U1}.

\section{Conclusions} 

In conclusion, we have presented the resistivity, magnetization, heat capacity and transverse field (TF) muon spin rotation ($\mu$SR) measurements in the normal and  the superconducting state of ThFeAsN, which has a tetragonal layered crystal structure.  Our resistivity and magnetization measurements confirmed the bulk superconductivity with T$_c $ = 30 K. From the TF $\mu$SR we have determined the muon depolarization rate in FC mode associated with the vortex-lattice.  The temperature dependence of $\sigma_{sc}$  fits  better to two-gap model either isotropic $s$+$s$-wave or $s$+$d$-wave than a single gap isotropic $s-$wave, anisotropic $s$-wave  or $d$-wave models. Our $\mu$SR analysis alone cannot distinguished between the $s$+$s$- and $s$+$d$-wave models,  but combining the results of field dependent heat capacity, we conclude that the $s$+$d$-wave model is the best to explain observed behavior of $\sigma_{sc}(T)$ and $\gamma$(H). Further, the value (for the larger gap) of 2$\Delta_1(0)/k_{\mathrm{B}}T_{\mathrm{c}}$ = 4.29$\pm0.01$ obtained from the $s$+$s$- and $s$+$d$-wave gap models fit is larger than 3.53, expected for BCS superconductors,  indicating the presence of strong coupling superconductivity in ThFeAsN. Moreover, two superconducting gaps have also been observed in the Fe-based families of superconductors and hence our observation of two gaps is in agreement with the general trend observed in Fe-based superconductors. Further confirmation of the presence of two gaps in ThFeAsN would require angle-resolved photoemission spectroscopy (ARPES) study on single crystals of ThFeAsN. The present results will help to develop a realistic theoretical model to understand the origin of superconductivity in ThFeAsN. 

\section*{ACKNOWLEDGEMENT}

DTA and HL would like to thank  the Royal Society of London for the UK-China Newton funding. DTA and ADH would like to thank CMPC-STFC, grant number CMPC-09108, for financial support. AB would like to acknowledge DST India, for Inspire Faculty Research Grant, and ISIS-STFC for funding support. The work at IOP, CAS was supported by the National Natural Science Foundation of China (NSFC-11374011 and 11611130165), the Strategic Priority Research Program (B) of the Chinese Academy of Sciences (XDB07020300). HL would like to thank the support from the Youth Innovation Promotion Association of CAS (No. 2016004).

\end{document}